\magnification=\magstep1
\baselineskip=16pt
\def\ni{\noindent}
\def\bs{\bigskip}
\def\eq{\eqno}
\input epsf


\centerline{{\bf  NEUTRON STARS: FORMATION AND
STRUCTURE}\footnote{$^{*)}$}{Lectures given at the XXV Mazurian
Lakes School of Physics, Piaski, August 1997.}}

\bs
\bs
\bs

\centerline{Marek Kutschera}

\bs

\centerline{H. Niewodnicza\'nski Institute of Nuclear Physics}

\centerline{ul. Radzikowskiego 152, 31-342 Krak\'ow, Poland}

\centerline{and}

\centerline{Institute of Physics, Jagellonian University}

\centerline{ul. Reymonta 4, 30-059 Krak\'ow, Poland}

\bs\bs\bs

\ni {\bf Abstract:}

A short introduction is given to astrophysics of neutron stars
and to physics of dense matter in neutron stars. Observed properties
of
astrophysical objects containing neutron stars are discussed. 
Current scenarios regarding formation and evolution
of neutron stars in those objects are presented. Physical
principles governing the internal
structure of neutron stars are considered with special emphasis on
the possible spin ordering in the neutron star matter.

\ni PACS: 21.65.+f,97.60.Jd

\bs\bs\bs
 
\ni {\bf 1. INTRODUCTION}

\bs

The existence of compact stars stabilized by the neutron pressure
was predicted soon after discovery of the neutron [1].
The idea of neutron stars is thus more than 60 years old. 
The first observation was made
only 30 years later in 1967. The history of the idea and
of actual discovery of neutron stars can be found in Ref.[1]. 

The first radio pulsar was discovered by Bell and
Hewish in November 1967 [1]. Soon after the discovery radio pulsars were
identified with rotating magnetized neutron stars [1]. Since then a
number of astrophysical objects involving neutron stars have
been observed both in radio and X-ray bands. These include
millisecond pulsars, binary radio 
pulsars, high-mass X-ray binaries and low-mass X-ray binaries.
Recently, an isolated nonpulsating neutron star which radiates
thermally was observed by the Hubble Space Telescope [2].
The star is  less than 130 pc from Earth.

The aim of these lectures is to give an introduction to both
astrophysics of neutron stars and physics of neutron star structure.
In Sect.2 a brief review of the astrophysical objects
which contain neutron stars is given. I will shortly describe
observed properties of these objects. In Sect.3 current scenarios of
formation of neutron stars in these objects are presented. In
Sect.4 I will outline some basic ideas regarding the
neutron star
structure. An emphasis will be given to the behaviour of proton
component of the neutron star matter. In particular,  a possible
spin ordering in the neutron star core and its implications for
neutron stars are discussed.

\bs

\ni {\bf 2. ASTROPHYSICS OF NEUTRON STARS}

Astrophysical objects in hydrostatic equilibrium fall in four
different categories according to the physical nature of
pressure which supports them against gravity. The source of
pressure in normal stars and in white dwarfs is the kinetic
energy of, respectively, a thermal plasma and a degenerate electron
gas. In planets and neutron stars the pressure is due to the
interaction energy of, respectively, a condensed atomic matter and
a condensed nucleon (hadronic) matter. 

Here we are interested
mostly in neutron stars. From the fact that hadronic pressure
becomes relevant only 
at a sufficiently high density when the nucleons start to feel their
mutual repulsion we can infer the density and the size of neutron stars.
The repulsive core of nucleon-nucleon interactions start to dominate at
densities exceeding the saturation density of nuclear matter,
$n>n_0=0.17 fm^{-3}$ (or mass density
$\rho>\rho_0=n_0(m_P-B/c^2) \approx 2.7\times 10^{14}g/cm^3$). 
The radius of a $1M_{\odot}$ star of density $\rho_0$ is $R\approx
12 km$. Hence neutron stars, objects stabilized by the nucleon (hadron)
pressure, have densities of the order of nuclear matter density
and radii of the order of $10 km$. 

Neutron stars, discovered initially as radio pulsars, are
responsible for a number of astrophysical phenomena. Below we
discuss in some detail astrophysical objects containing neutron stars.

\bs

\ni {\bf 2.1. Radio pulsars}

Radio pulsars are rapidly rotating strongly magnetized neutron
stars. Their rotation period, $P$, is in the range $0.033s \le P
\le 4.0 s$. Pulsars emit radiation at the expense of the rotational
energy. As a result, they spin down. The spin-down rate ${\dot
P}$ is measured to be $\sim 10^{-12}-10^{-16}ss^{-1}$.
Measurements of the period, $P$, and its derivative, ${\dot P}$,
allow us to obtain the magnetic field of the pulsar, B, assuming
that the spin-down is due to the dipole magnetic braking. A simple formula
for the pulsar "luminosity" is 

$$L_{pulsar}={2 R^6 \over 3c^3} ({2 \pi \over P})^4 B^2=-{\dot
E_{rot}}, \eq(1)$$

\ni where the pulsar rotational energy is $E_{rot}=I(2\pi/P)^2/2$.
From Eq.(1) we find that the magnetic field is $B=3.2\times10^{19}\sqrt{P{\dot
P}}$, with $P$ in seconds. In this estimate, typical values of
the moment of inertia, 
$I=10^{45}gcm^2$, and the 
radius, $R=10km$, are used. Pulsar magnetic fields  are found to
be $B\sim 10^{11}-10^{13} G$. 

The total radio luminosity of pulsars is generally  $\sim 10^{26} - 10^{30}
erg/s$. These values are some two to six orders of magnitude
below the rate ${\dot E_{rot}}$ of rotational energy
dissipation. Pulsars obviously emit energy at other wavelengths.

One of the best studied pulsars is the Crab pulsar, PSR 0531+21,
with the period $P=0.033s$. The neutron
star (Crab pulsar) and the Crab nebula in which the pulsar is
embedded are remnants of the
supernova in the year 1054. The Crab pulsar is thus about $940$ years
old. It is the youngest known pulsar. The Crab pulsar is 
observed optically as a faint star near the center of the Crab
nebula. The  pulsar radiates also in X- and
$\gamma$-rays. In all bands, at radio frequences, in visible
light, in X-rays and in $\gamma$-rays the signal displays the
same pulse profile, Fig.1. According to current theories of
pulsar emission, the pulsating radiation at all wavelengths is
produced by the same magnetospheric activity which generates the
radio signal. This means that the visible light is not
emitted from the surface of the neutron star but rather from
an extended magnetosphere. Optical observations of the Crab pulsar
do not allow us to constrain the radius of the star. 

\bs

\ni  {\it 2.1.1. Millisecond pulsars}

There exists a clearly distinct category of radio pulsars,
millisecond pulsars, with short periods in the range $1.56 ms \le P
\le 100 ms$ and very small period derivatives, ${\dot P} \sim
10^{-17} - 10^{-21}$. The fact that 
corresponding magnetic fields are much weaker than for ordinary
pulsars, $B \sim 10^8-10^{10} G$, strongly suggests that
millisecond pulsars should be treated as a different class of
radio pulsars. Almost one half of all millisecond pulsars are
members of binary systems, which is  another distinct feature of
these objects. The fastest known pulsar, PSR 1937+21, has the
period $P=1.56 ms$ and ${\dot P}=10^{-19}$. Some researchers
[3] postulate that millisecond pulsars with lowest
magnetic fields, $B \sim 10^8 G$, are of different origin than
those with higher fields. 

\bs

\ni {\bf 2.2. X-ray binaries}

X-ray observations made by satellite experiments have revealed
the existence of binary systems composed of the neutron star with the
companion which is a normal star. There are two classes of such
binary stars, high mass X-ray binaries and low mass X-ray
binaries. In both cases, the neutron star  is accreting matter from
the companion, which results in X-ray emission. However the
pattern of the X-ray emission is different for both classes.

\bs

\ni {\it 2.2.1. High mass X-ray binaries}

The X-ray flux of high-mass X-ray binaries displays a periodic
modulation, hence these objects are 
also called X-ray pulsars. The companion of the neutron star is
usually a massive O or B 
star with the mass $\sim 10 M_{\odot} - 40 M_{\odot}$. The
neutron star is strongly magnetized with the magnetic field 
typical for normal radio pulsars, $B \sim 10^{12} -
10^{13} G$. Matter accreted onto the neutron star moves along
the magnetic field lines and hits the neutron star surface in
the vicinity of the magnetic pole. In this way, a hot spot
emitting X-rays is
formed around the magnetic pole. Rotation of the neutron star around
an axis inclined with respect to the magnetic axis produces the
pulsar effect. The rotation period of X-ray pulsars 
ranges from $P \sim
0.1s$ to a fraction of an hour. The slowest X-ray pulsar has the
period $P \sim 1400 s$ 
[4]. Since heavy stars live short the age of high mass
X-ray binaries is estimated to be less than $10^7 y$.
Recent review of X-ray pulsars is given in Ref.[5].

One should note that for some accreting X-ray pulsars the
rotation period decreases, ${\dot P}<0$. Apparently, these
pulsars are spinning up due to accretion of matter.

\bs

\ni {\it 2.2.2. Low mass X-ray binaries}

The companion of the neutron star in low mass X-ray binaries is
a faint star 
with the mass $M \le 1.2 M_{\odot}$. The X-ray emission displays
bursts which repeat in several hours. Low mass
X-ray binaries are 
also named X-ray bursters. The X-ray bursts are thought to be
produced by a thermonuclear flash of the accreted nuclear fuel.
Accreted matter spreads over the surface of the neutron star.
This occurs because the magnetic field is low, $B \le 10^{10} G$.
Continuous accretion from the companion increases the density
and temperature at the base of the fuel shell. When the
condition of instability is reached, a nuclear flash occurs. A new
shell of fuel is then accreted and the cycle repeats. 
The  X-ray busters are rather old objects,
with the age more than $10^8$ years.

\bs

\ni {\bf 2.3. Binary radio pulsars}

An important class of neutron star binaries are systems where
the neutron star is detected as a radio pulsar. Such systems are
binaries with no mass transfer.
The best studied binary radio pulsar
is the Hulse and Taylor pulsar, PSR 1913+16 [6]. 
Generally, one
can distinguish two subclasses of these objects, close systems
with a
compact companion, PSR 1913+16
subclass, and loose 
systems with the companion being low-mass faint star, PSR 1953+29 subclass.
In the former case the
orbit is tight and eccentric with the orbital period $P_{orb}
\sim 1 day$. The compact 
companion is a neutron star or a massive white dwarf
$(M \approx 1.4M_{\odot})$. In the latter case the system
is more loose with the orbital period  $P_{orb} \sim 1 year$ and
with circular orbit.
The companion is a low mass helium white dwarf
with mass $M \le 0.4 M_{\odot}$. At present six double neutron
stars are known. 

Observations of double neutron stars are very important. Shrinking of
the orbit in the pulsar PSR 1913+16 is the best indication
available of gravitational radiation. Also, precise
determination of masses of both neutron stars is possible for
double neutron stars. This is crucial for constraining models of
neutron star interior.

Binary radio pulsars play also important role in our
understanding of the evolution of neutron stars in binaries, as is
discussed below.

\bs

\ni {\bf 2.4. Thermally radiating single neutron stars}

Astrophysicists expect that there are some $\sim 10^8$ neutron stars in
our Galaxy. These are remnants of previous generations of
massive stars which produced all the metals (i.e. $Z > 2$ elements)
observed at present. Many of these stars should be hot enough
and close enough to be observable as X-ray sources. Recently, the
first such star was observed both in X-ray band [7] and  
in UV and visible band by the Hubble Space
Telescope [2]. The neutron star is hot and it was
first observed as a strong X-ray source by the ROSAT satellite.
The star emits thermal radiation and is located about 130 pc
from Earth. This allows for the first time to directly constrain
the radius of a neutron star to be $R \le 14 km$ [2].

\bs

\ni {\it 2.4.1. Invisible neutron stars}

Nonrotating and nonaccreting neutron stars which have cooled
down are virtually undetectable by conventional means. Such
stars contribute to the dark matter of the Galaxy in the form of
massive compact objects. Recently, gravitational microlensing
experiments have detected a population of such objects in the
galactic halo [8]. Future microlensing observations are crucial for
determination of the contribution of invisible massive baryonic
objects to the total mass of the Galaxy. In particular they
should answer the question if there is a halo of neutron stars
around our Galaxy. Such a halo composed of stars which obtained
sufficiently high initial velocity at birth would explain
scarcity of local thermally radiating neutron stars.

\bs

\ni {\bf 2.5. Physical parameters of neutron stars}

Observations of neutron stars in all astrophysical objects discussed above
allow us to determine some of their physical parameters. The
relevant parameters are
the mass,  the radius, the temperature, the age and the magnetic
field. Generally, simultaneous measurement of all above
parameters for a given neutron star is not possible. 

\bs

\ni {\it 2.5.1. Neutron star masses}

Presently, masses of about 20 neutron stars in binary systems are
determined. Among them are six double neutron star binaries, PSR
B1913+16 [9], PSR B1534+12 [10], PSR B2303+46 [11], PSR B2127+11C
[12], PSR B1820-11 [13] and PSR J1518+4904 [14]. For three of them,
PSR B1913+16, PSR B1534+12 and PSR B2127+11C,
precise measurements of mass of both neutron stars are now 
available. The masses are found to be $M_1=1.44 
M_{\odot}$ and $M_2=1.39 M_{\odot}$ (PSR B1913+16), $M_1=1.34
M_{\odot}$ and $M_2=1.34M_{\odot}$ (PSR B1534+12) and $M_1=1.35
M_{\odot}$ and $M_2=1.36 M_{\odot}$ (PSR B2127+11C).
All masses lie in a rather narrow interval,
$1.3 M_{\odot}<M<1.5 M_{\odot}$. Masses of neutron stars in
X-ray pulsars are also consistent with these values although
are measured less accurately. The measured masses of neutron
stars apparently do not exceed the maximum mass which is about
$1.5M_{\odot}$.  We shall discuss possible implications of this
upper limit in the last section.

\bs

\ni {\it 2.5.2. Neutron star radii}

The first observation of thermally radiating nonpulsating neutron
star RX J185635-3754 [2,7] constrains the radius of the star to be $R< 14 km$.
Generally, radii of neutron stars are not directly observable. One can
infer, however, some plausible values from model calculations of
X-ray bursters  which are of the order of $10 km$. 

\break
\vfill

\ni {\it 2.5.3. Surface temperature of neutron stars}

The X-ray and optical spectrum of the neutron star RX
J185635-3754 corresponds to  the black body
with $k_BT\approx 60 eV$. Also,
the X-ray flux from about 14 pulsars has been  detected [15]. The
spectrum of photons is more difficult to obtain. If measured, it
is often not consistent with the thermal emission but is rather
dominated by a hard component due to the magnetospheric activity. Only
for four pulsars, PSR 0833-45 (Vela), PSR 0630+18 (Geminga), PSR
0656+14 and PSR 1055-52, softer blackbody component
corresponding to surface thermal emission is determined. 

\bs

\ni {\it 2.5.4. The age of pulsars}

Pulsar ages are estimated by measuring their spin-down rates. 
Pulsars spin down due to conversion of rotational 
energy into radiation. A simple spin-down relation is assumed,
${\dot \nu}=K\nu^n$, where $\nu$ is the rotation frequency, and $n$ is
the braking index. The constant $K$ 
for magnetic braking is proportional to $d^2/I$, where $d$ is
the magnetic moment of the star and $I$ is the moment of
inertia. If the energy loss is through radiation from a
dipolar magnetic field, the braking index is $n=3$. The
spin-down age of pulsar is then $\tau=-\nu/2{\dot \nu}$. 

The spin-down age with $n=3$ is commonly used for pulsars. Its
applicability, however, is questioned by
recent measurement of the braking index of the Vela
pulsar [16], which gives $n=1.4\pm0.2$. This value
implies that previous estimate of the age of Vela pulsar should
increase by a factor $\sim 3$. 

Typical values of the spin-down age for radio pulsars are $10^7
y$. For millisecond pulsars the spin-down age exceeds $
10^9 y$.

\bs
\ni {\it 2.5.5. Magnetic fields of neutron stars}

Magnetic fields of radio pulsars are inferred from the spin-down
relation assuming dipolar magnetic field. A striking feature is
the bimodal distribution of pulsar magnetic fields. Usual
pulsars have strong magnetic field, $B\sim 10^{11}-10^{13} G$,
whereas millisecond pulsars possess much weaker  fields,
$B\sim 10^8-10^{10} G$. For some X-ray pulsars the magnetic
field is measured directly, by observation of absorption
features interpreted as cyclotron lines [17]. The values are in
the range found for normal pulsars. It should be noted that 
neutron stars in the X-ray bursters have, if any, weaker
fields, $B<10^{10} G$.

The magnetic field of neutron stars could also serve as a
probe of the neutron star equation of state (EOS) if its
presence is determined by 
the properties of dense matter.
The bimodal distribution of pulsar magnetic fields
strongly suggests existence of a magnetic phase transition in
the neutron
star matter [17]. There is a possibility, however, that some
component of the magnetic field of a neutron star is
inherited from the progenitor.

The magnetic phase transition in the neutron star matter is
discussed in Sect.4.

\bs

\ni{\bf 3. FORMATION OF NEUTRON STARS}

Neutron stars are the final product of stellar evolution. The
evolution of single stars is significantly different than the
evolution of stars in binaries. The formation
scenario of neutron stars in binary systems reflects this fact. We
briefly discuss below both cases. 

\bs

\ni {\bf 3.1. Evolution of single stars}

Life of single stars with the mass $M \ge 8M_{\odot}$ is
terminated by
the supernova explosion.
The explosion is a result of the gravitational collapse of the
star's iron core whose mass has exceeded the Chandrasekhar mass,
$M_{Ch} \approx 1.5 M_{\odot}$, the maximum mass which can be
supported by pressure of the electron gas. The
collapse is halted when the density increases sufficiently for
the nucleons to feel the repulsive core of their mutual
interaction. The nuclear interactions produce very steep
pressure gradient which is able to stop the collapse. Rapid
release of the
gravitational energy, which is carried mostly by neutrinos, causes the
violent explosion of outer layers of the star. The total amount
of the energy released in the gravitational collapse of the
stellar core is $\sim
10^{53} erg$. Only $\sim 1\%$ of this energy is needed to
expel the mantle and envelope of the collapsing star.
The basic scenario of the Type II supernova explosion was
spectacularly confirmed by detection by the Kamiokande
experiment of the neutrino burst
associated with the supernova SN 1987A. The energy of neutrinos
and the distance to the Large Magellanic Cloud, where SN 1987A
occurred, allows to estimate the total energy carried by
neutrinos to be $\sim 10^{53} erg$.

\bs

\ni {\bf 3.2. Evolution of binaries}

Evolution of stars in close binaries in which stars can exchange
mass proceeds differently than for single stars.
Evolution of binaries relevant to neutron star formation depends
on the initial mass and separation of the primary and secondary
star. Below we discuss the origin of high mass X-ray binaries
and low mass X-ray binaries and their evolutionary links to
binary and millisecond pulsars.

\bs

\ni {\it 3.2.1. Formation of high mass X-ray binaries}

The progenitor of the high mass X-ray binary is a binary system
with the primary star heavier than the O or B star present in
such X-ray binaries, $M_1>M_2$. This star evolves quickly and
explodes as a 
supernova. In general, the binary system can be disrupted by the
supernova explosion. However, if the mass ejected from the
system is less than the mass left, 

$$M_{ejected}<M_2+M_{NS} \eq(2)$$

\ni where $M_{NS}$ is the mass of the neutron star, the binary
binding can survive [18]. Also, a suitable natal kick imparted to
the neutron star at birth in asymmetric supernova explosion, can save the
binary from disruption. 
For the produced system to become an X-ray binary sufficient
accretion onto  the neutron star is required. Since
the newly formed neutron star has a
strong magnetic field, it becomes an X-ray pulsar when it accretes
matter from the normal companion. Such an interacting binary
system can be formed for a range of masses and orbital
separations of the primary and secondary stars of the progenitor.

\bs

\ni {\it 3.2.2. Origin of double neutron stars}

There exist scenarios linking the formation of the binary
pulsars of PSR 1913+16 class to the evolution of high mass X-ray
binaries. In these models the evolution proceeds through
the common envelope phase.
The companion of the neutron star in the high mass X-ray binary,
which is primary as far as 
the mass is concerned, evolves quite fast. When it approaches
the red giant phase, its expanding
hydrogen envelope can engulf the neutron star. The binary system
enters a hypothetical period of common envelope evolution.
The neutron star is expected to spiral inward due to friction,
getting closer to the helium core. In this process, the envelope
which is weakly bound is thought to become expelled. As a
result, a binary system of the neutron star and the helium star,
is formed. The mass of the helium star is about $3 M_{\odot}$.
Such a helium star after some time undergoes core collapse and
explodes as a supernova which leaves a remnant neutron star.
Detailed considerations suggest [19] that only in an asymmetric
supernova explosion
a binary system of two neutron stars can be formed.

The binary pulsar PSR 1913+16, which is a double neutron star,
has likely followed such an evolutionary track. Similar
evolutionary considerations are valid for other pairs of neutron
stars. These systems are observed as binary radio pulsars with a
neutron star companion. An important question is which of the
two neutron stars is the radio pulsar. The fact, that the pulsar
in PSR 1913+16 has period $59 ms$ and low magnetic field $B\sim
2\times10^{10}G$ suggests that it is a recycled pulsar with the age
$\sim 10^8 y$. Newly formed neutron star is expected to have much
stronger magnetic field, $B\sim 10^{12} G$, and to live as a
radio pulsar only $\sim 10^7$ years. Hence it is more likely that we
observe as a radio pulsar the older neutron star in this binary.

The orbits of binary pulsars with neutron star companions shrink
due to gravitational 
radiation. The two neutron stars will collide/merge in some
$10^9 y$. The collision/merger is unavoidable. In such a violent
event a lot of energy will be liberated in a rather short time
interval and the collision can manifest itself as a powerful
explosion. Many researchers consider such a neutron star
collision the best candidate for gamma-ray bursters.

\bs

\ni {\it 3.2.3. Formation of low-mass X-ray binaries}

The origin of low mass X-ray binaries is more difficult to
explain [20]. Most of evolutionary models base on common envelope
evolution prior to the supernova explosion. The major problem is
the asymmetry of masses of assumed 
progenitors of these binaries. The  masses of donor stars, which
are companions of neutron stars, are low, $M \le
1.2 M_{\odot}$, and these stars are unevolved. The mass of the primary of
the progenitor must be high enough in order to produce a neutron star.
It can exceed the secondary mass, which is the donor mass
of low mass X-ray binary by a factor $ \ge 20$. The evolutionary
sequence leading to the low mass X-ray binary consists of
several steps. First, a common envelope forms, due to small time
scale of nuclear evolution of the massive primary. During this
phase the hydrogen envelope is dispersed and the binary orbit is
shrunk. Otherwise, if
the primary explodes as a supernova before the envelope is
expelled the system becomes disrupted. After the common envelope
phase the binary system with a helium star emerges. The chance
the system  survives the explosion of the helium 
star is increased, in particular the
asymmetric explosion imparting a proper kick to the neutron star
can help.  

An alternative scenario of formation of low mass X-ray binaries
is the accretion-induced collapse of the white dwarf. In this
model, the Chandrasekhar mass white dwarf collapses to the
neutron star when some matter is accreted onto it. 
The progenitor binary consists of two low mass stars, with
the heavier one becoming the carbon white dwarf. It remains an
open question if the accretion induced collapse of the white
dwarf occurs in nature. One should remember that the carbon
white dwarf contains a lot of nuclear fuel which can undergo
thermonuclear flash, that completely disrupts the star. Such
an explosive carbon burning of a white dwarf in a binary is a
successful model of the Type I supernovae. No remnant star is
left in such an explosion. 

Those binaries which sustain low accretion rates for a very long
time are observed as low mass X-ray binaries. Magnetic fields of
neutron stars in these binaries seem to be low, $B\le 10^{10}G$.
One popular explanation is that the magnetic field, initially of
the order $10^{12}G$ has decayed due to accretion. 

\bs

\ni {\it 3.2.4. Origin of millisecond pulsars}

Accretion in low mass X-ray binaries lasts for such a long time
that a total mass of a few tenths of solar mass, $M_{tot} \sim 0.1
M_{\odot}$, is transferred 
from the low mass star to the neutron star. 
When the accretion ceases the donor is thus a helium white dwarf
typically with $M \le 0.4 M_{\odot}$. The neutron star which has
accreted
$\sim 0.1 M_{\odot}$ is spinning rapidly due to acquiring
additional angular momentum transferred by the accreted matter.
 The
magnetic field of the neutron star is now weaker but the
neutron star can manifest itself as the millisecond pulsar
in the binary system. The neutron star becomes a recycled
pulsar. Thus low mass X-ray binaries seem to be
progenitors of binary millisecond pulsars [21].  

There are several examples of binary millisecond pulsars with
very low mass companion. The pulsar PSR 1744-24A with $P=11.6 ms$ has
the companion of mass $M_c=0.1 M_{\odot}$ and the $1.61 ms$ pulsar PSR
1957+20 has the one with $M_c=0.02 M_{\odot}$. Apparently,
millisecond pulsars in these binaries  evaporate their
companion stars. The binding energy of low mass stars is
$E_b \le 10^{48} (M_c/M_{\odot})^2 erg$ whereas the rotational
energy of the pulsar is $E_{rot}=I\omega^2/2 \sim
10^{52}P_{ms}^{-2} erg$. The pulsar can evaporate the companion
if $E_b \le gE_{rot}$, where the factor $g$ is the efficiency of
evaporation.
Complete evaporation of the low mass companion by the
millisecond pulsar is  currently the preferred scenario of
the formation of single millisecond pulsars [22].
\bs

\ni {\bf 4. STRUCTURE OF NEUTRON STARS}

The internal structure of neutron stars is determined by
properties of matter at supernuclear densities which are subject to 
considerable uncertainty.  At densities close to the saturation
density the composition of neutron star matter can be determined
unambiguously by simple extrapolation from laboratory data.
With increasing density, one needs to employ models of nucleon
interactions which introduce some uncertainty. The
higher the density the less reliable is the  
equation of state. In particular, composition of dense matter
can change and new degrees of freedom, such as hyperons,
condensed meson fields and
quarks should be taken into account.

There exist recent reviews of physics of neutron star matter
[23] and the equation of state [24]. I have selected some topics
relevant to internal properties of neutron
stars in the density range where most likely the neutron star matter
is composed of nucleons and leptons only. I will focus on
possible localization of protons and spin ordering in 
the neutron star matter. Existence of such a phase can influence
magnetic fields of neutron stars.

\bs

\ni {\bf 4.1. Crust}

We shall focus on structure of physically relevant neutron star of mass
$\sim 1.4 M_{\odot}$. The surface of the star is defined at the radius where 
pressure vanishes. Just below the surface the pressure is dominated by the
degenerate electrons of the solid crust. The crust is similar to matter 
from which white dwarfs are made of. It is the lattice of bare nuclei
immersed in the electron sea. With increasing density, deeper in the star,
a layer occurs where the mean matter density becomes equal to the nuclear
matter density $n_0=0.17 fm^{-3}$. This is the boundary of the neutron star
core. The core is composed of nucleon liquid of densities
exceeding the saturation density, $n>n_0$. 

\bs
\ni {\bf 4.2. Neutron star core}

The material, from which the neutron star core
is built of, is condensed hadronic matter, essentially in the
ground state. Such matter is named cold catalysed matter. Just
below the crust, the core matter is composed 
mainly of neutrons with some admixture of protons, electrons and
muons.
This matter is condensed since the Fermi energy of
electrons is much higher than the thermal energy, $E_F>>k_BT$.
Thermal energy of particles in the neutron star is below $1
MeV$, except of first few minutes after formation.
The temperature of a newly
born neutron star is $k_BT \sim 10 MeV$. The star cools to $k_BT
\sim 1 MeV$ in a few minutes.
The electron Fermi energy in the core is $E_F \sim 100 MeV$.

Weak interactions ensure that the neutron star matter relaxes to
$\beta$-equilibrium with neutron, proton and electron chemical
potentials satisfying the condition

$$ \mu_N=\mu_P+\mu_e. \eq(3)$$

\ni When the electron chemical potential exceeds the muon rest
mass, $\mu_e \ge m_{\mu}=106 MeV$, muons appear in the matter
with the chemical potential $\mu_{\mu}=\mu_e$.

The neutron star matter is locally charge neutral, with the
lepton (electrons + muons) density equal to the proton density,

$$ n_e+n_{\mu}=n_P. \eq(4)$$

\bs

\ni {\it 4.2.1. Neutron star matter at density $n_0$}

Properties of neutron star matter of saturation density can be
inferred from empirical parameters of nuclear matter which are obtained
from nuclear mass formulae. The empirical value of nuclear
symmetry energy, $E_s=31\pm 4 MeV$ allows one to fix the proton
fraction by using Eq.(3) and (4). The proton fraction,
$x=n_P/n$, of $\beta$-stable nucleon matter of saturation density is
$ x(n_0)\approx 0.05$.

 The surface layer of the neutron star core, of density
$n_0\approx 0.17 fm^{-3}$, is the only part of the core whose
composition is determined quasi-empirically. The value
of the proton fraction, $\sim 5\%$, is known with accuracy determined
entirely by the empirical error of the nuclear symmetry energy.

The composition of $\beta$-stable matter of density $n_0$ is
found by minimization of the energy of charge neutral nucleon
matter with leptonic contribution with respect to proton
fraction. The energy per baryon is

$$ E(x)=E_{nuc}(x)+E_{lepton}(x), \eq(5)$$

\ni where the nucleon contribution is

$$ E_{nuc}(x)=E_0+E_s(2x-1)^2. \eq(6)$$

\ni In the last formula $E_0=-16 MeV$ is the saturation energy
of symmetric nuclear matter. The formula (6) is the $A\to
\infty$ limit of the semiempirical mass formula. 

The lepton contribution consists of
both electron and muon parts, $ E_{lepton}=E_e + E_{\mu}$. Muons
appear in the system when the electron chemical potential
exceeds the muon rest mass, $\mu_e \ge 106 MeV$, which happens
at $n_e \ge 0.005 fm^{-3}$. The lepton part can be calculated
exactly. At densities where $n_{\mu}=0$, the electron part is
$E_e=250 (n_B/n_0)^{1/3}x^{4/3}$ $MeV$, and in the high density limit
with $n_e=n_{\mu}$, $E_{lepton} \to 200 (n_B/n_0)^{1/3}x^{4/3}
MeV$.

The minimum of $E(x)$ occurs at $x\approx 0.05$. Hence the
macroscopic nuclear matter of density $n_0$ consists of protons,
neutrons, electrons and muons with densities $n_P=0.05n_0,
n_N=0.95n_0, n_e=0.99n_P$ and $n_{\mu}=0.01n_P$. At the minimum
the electron chemical potential is
$\mu_e=\mu_{\mu}=\mu_N-\mu_P=110 MeV$. The energy per baryon is
$E=8MeV$. The properties of neutron star matter of density
$n_0$ are fully determined by the empirical parameters of
nuclear matter.

\break
\vfill

\ni {\it 4.2.2. Neutron star matter at higher densities}

For densities $n>n_0$ no direct empirical data relevant to the
neutron star matter are available. In order to study its
properties one must resort to models of nucleon matter. There
exist a variety of realistic models of nucleon matter. One class
is based on effective interactions, such as Skyrme forces [25] or
Myers and Swiatecki interactions [26]. These are generally developed
by fitting properties of finite nuclei. Another class is based
on phenomenological potentials, such as e.g. Reid soft core
potential [27], Urbana $v_{14}$ potential [28] or Argonne
$v_{14}$ potential [29], and employs
variational many-body theory to calculate the equation of state
of dense matter [30]. Phenomenological potentials are constrained to
fit the nucleon-nucleon scattering and the properties of the
deuteron. The models provide energy per particle as a function
of proton and neutron density. This allows us to calculate
chemical potentials of protons and neutrons and to construct the EOS of
$\beta$-stable neutron star matter. 

Despite the fact that the realistic 
models predict rather different equations of state, there is one
qualitative feature which is similar: the proton fraction of the
$\beta$-stable neutron star matter, $x(n_B)$ (Fig.2). Generally,
realistic models predict
that the proton fraction is low, $x(n) \le 0.1$, and, as a
function of baryon density $n_B$, decreases with $n_B$. At high
enough densities protons eventually disappear from
$\beta$-stable neutron star matter. The density $n_v$ at which
the proton fraction vanishes is model dependent. From Fig.2 one
can see that $n_v \sim 0.5 fm^{-3}$ for some parametrization of
Skyrme forces and  $n_v \approx 1.0fm^{-3}$
for the UV14+TNI
interactions. For Myers and
Swiatecki interactions $x(n_B)$ drops slowly with density.

Generally, realistic models of nucleon matter lead to a qualitative
prediction that at high densities proton content of neutron star
matter decreases and eventually vanishes. At densities slightly
lower than $n_v$ protons are impurities in the neutron star
matter. In the following we shall discuss properties of such a
strongly asymmetric nuclear matter which have important
consequences for neutron stars. 

\bs

\ni {\it 4.2.3. Why do protons disappear?}

It seems strange at first sight that at densities $n_B>n_v$ pure neutron matter
is $\beta$-stable. This behaviour can be understood as a result
of nucleon-nucleon interactions used in realistic models of the nucleon
matter. 

The fact that proton fraction vanishes at density $n_v$ means
that the minimum of the energy per baryon (5) occurs at this
density for the first time at $x=0$. This can be traced back to
the behaviour of the 
nucleon contribution, $E_{nuc}$ [31].
In analogy with Eq.(6) we write
the nucleon energy in the form

$$ E_{nuc}(n_B,x)=E_{nuc}(n_B, {1 \over 2}) + E_s(n_B) (2x-1)^2,
\eq(7)$$

\ni where $E_{nuc}(n_B, {1 \over 2})$ is the energy of symmetric
nuclear matter at density $n_B$ and $E_s(n_B)$ is the symmetry
energy at $n_B$. As long as the symmetry energy is positive,
$E_s(n_B)>0$, the minimum of the total energy
$ E_{nuc}+E_{lepton} $ occurs at some finite proton fraction $ x \ne 0$. However,
when $E_s=0$, the minimum occurs at $x=0$. This means that the
density $n_v$ corresponds to vanishing of the symmetry energy,
$E_s(n_v)=0$. At higher densities, $n_B>n_v$, the symmetry energy
is negative, $E_s(n_B)<0$. 

Change of sign of the symmetry energy is the immediate cause of
disappearance of protons. With $E_s<0$, Eq.(5) shows the energy
per particle of pure neutron matter ($x=0$) is lower than for any
$x>0$. It is thus energetically favourable for the system to get
rid of all protons and negatively charged leptons [32].

\bs

\ni {\it 4.2.4. Microscopic origin of proton disappearance}
 
The $\beta$-stability condition (3) at the density $n_v$ becomes

$$ \mu_N(n_v,x=0)-\mu_P(n_v,x=0)=0 \eq(8)$$

\ni as the electron chemical potential also vanishes. The
formula (8) allows us to understand the microscopic origin of
the phenomenon of proton disappearance. 

The chemical potential of neutrons is the neutron Fermi energy
which has both kinetic and interaction contributions,

$$ \mu_N = E_{kin}(k_N^F)+ E_{int}^{NN}, \eq(9)$$

\ni where $k_N^F$ is the neutron Fermi momentum and $E_{int}^{NN}$
is the interaction energy of the neutron at the Fermi surface
with other neutrons. The proton chemical is

$$ \mu_P = E_{int}^{PN}, \eq(10)$$

\ni where  $E_{int}^{PN}$ is the interaction energy of zero
momentum proton with surrounding neutrons. Formally, this is the Fermi
energy of an empty proton Fermi sea and the proton Fermi momentum is $k_P^F=0$. 
 
From Eq.(8) we see that

$$  E_{int}^{PN}> E_{int}^{NN}. \eq(11)$$

\ni The fact the proton-neutron interaction energy is higher
than neutron-neutron interaction energy means that in the
underlying nucleon-nucleon potentials the proton-neutron
repulsion is stronger than the neutron-neutron one. For several
phenomenological potentials this really is the case. The central
potentials for the proton-neutron channel and the neutron-neutron one
can be written as $ v_{PN}=v_c-v_{\tau}$ and $v_{NN}=v_c+v_{\tau}$
where $v_c$ and $v_{\tau}$ are central and isospin
components of  general nucleon-nucleon potential

$$ v_{nucleon-nucleon}(r) = v_c(r) + v_{\tau} {\vec \tau}_1{\vec
\tau}_2+... \eq(12)$$ 

\ni where dots represent not displayed contributions (tensor,
spin, spin-orbit, etc.). 
The $v_{\tau}$ component for popular potentials is shown in
Fig.3. As one can see, $v_{\tau}<0$, and thus $v_{PN}>v_{NN}$.

\bs

\ni {\bf 4.3. Proton impurity in the neutron star matter}

Nuclear matter in neutron stars which contains only small
admixture of proton impurities displays some new features, not
encountered at higher proton concentrations. Below we show that
impurities likely become localized at high densities. This means
that nucleon matter is no longer a simple Fermi liquid. 
Particularly important for neutron stars is the  ferromagnetic
instability of neutron matter with localized protons.

The first observation is that the proton impurity couples to density
fluctuations [33]
of the neutron matter with the interaction Hamiltonian of the
deformation-potential form. To show this let us consider the
energy of a single proton in the uniform neutron matter of
density $n_N$, which can be written in the form [34]

$$ E_P({\vec k})= {k^2 \over 2m_*} + \mu_P(n_N). \eq(13) $$

\ni The single impurity case corresponds to ${{\vec k} \to 0}$.
The last term in this formula represents the interaction energy
of the proton impurity with the neutron matter. As shown above, this is
the proton chemical potential. 

Let us assume that there are small density oscillations in the
neutron matter,

$$ n({\vec r}, t)=n_N + \delta n({\vec r}, t).  \eq(14) $$

\ni The Hamiltonian of the proton impurity in the presence of
small density oscillations becomes

$$ H_P= - {\nabla^2 \over 2m_*} + \mu_P(n_N) + {\partial \mu_P
\over \partial n_N} \delta n({\vec r}, t). \eq(15) $$

\ni The last term represents the coupling of the proton impurity
to phonons of the neutron matter. After quantizing the
oscillations we obtain [33] the deformation-potential coupling
of Bardeen and Shockley [35]  for the polaron problem in solids.
This interaction is characterized by the deformation-potential
coupling constant 

$$ \sigma= {\partial \mu_P \over \partial n_N} n_N. \eq(16) $$

\ni The coupling vanishes for the density where the proton
chemical potential in pure neutron matter has the minimum [33].
This density is close to $n_0$. At higher densities $\sigma$
increases with density [33].

The proton impurity induces a virtual hole in the neutron medium
which moves with it. The hole can be interpreted as a phonon
cloud accompanying the proton. The proton impurity displays thus
a typical polaron behaviour [33]. 
This  affects the effective mass of the 
impurity which is now the nuclear polaron mass,
$m_{polaron}>m_*$. 

The regime of small values of the
deformation-potential coupling $\sigma$ corresponds to so called large
polaron [33]. With increasing $\sigma$ one enters the strong
coupling regime where eventually the polaron becomes localized
[33]. The localized polaron is also called small polaron.
Localization of the proton impurity, which is the nuclear
polaron, corresponds to freezing of the deformation of the
neutron background. This is a phase transition. The localization
is a selfconsistent trapping of the proton wave function in
the potential well corresponding to the hole induced by the
impurity in the neutron background. The localization density
$n_{loc}$ depends on the model of nucleon interactions. In Ref.[33] we
have found $n_{loc} \sim 3n_0$ for the UV14+TNI interactions.

\bs

\ni {\bf 4.4. Ferromagnetic instability of neutron star matter
with localized protons}

It is easy to see that a single localized proton induces a spin
excess of the neutron background provided any proton-neutron spin
interaction exists. The nucleon-nucleon interactions are of
course strongly spin dependent and thus there always exists some
proton-neutron spin interaction,

$$v_{spin}= v_{\sigma}(r) {\vec \sigma}_N{\vec \sigma}_P. \eq(17)$$
 
\ni Since this spin interaction is linear the proton spin can always
be arranged in such a way that the net effect is attractive. For
a fixed neutron spin, if $v_{\sigma}>0$($v_{\sigma}<0$) the
proton spin should be antiparallel (parallel) with respect to
the neutron spin. For a single proton, there is no Pauli
blocking and the proton spin can be changed with no increase of
the kinetic energy. It is thus obvious that the system will
relax to the spin ordered configuration with the lowest
energy [34]. Such a configuration will possess induced spin
excess of the neutron background in the vicinity of the proton
impurity [34]. 

At low proton fractions, the above analysis also applies [36].
The neutron star matter which contains localized protons 
exhibits ferromagnetic instability [37]. One can identify mechanism
by which the long-range order is produced [38]. This is the well
known RKKY mechanism operating e.g. in rare-earth alloys. The
basic physics is as follows. The 
spin excess of the neutron background induced by a given proton
extends outside the region where the proton wave function is
localized. In this way  a neighbouring localized proton is
polarized in the direction of the spin of the first proton, etc.
A long-range ferromagnetic order can propagate in the system. 

\bs
\ni {\it 4.4.1. Neutron stars with ferromagnetic core}

Neutron star matter with localized protons is magnetized [36].
The magnetization can be parametrized as [40]

$$M=\mu_{eff}n_P, \eq(18)$$

\ni where $\mu_{eff}$ is the effective magnetic moment of the
localized proton [36]. Typical values of magnetization are $M
\sim 10^{13}G$ [39]. This value 
is rather similar to typical magnetic fields of normal radio
pulsars. 

The contribution of the magnetized neutron star matter
to pulsar magnetic field was considered in  [40].  
If the proton localization density  is less than the central
density of the neutron star, $n_{loc}<n_c$, the star develops
the ferromagnetic core. Such a core produces a magnetic moment
of the neutron star.

\bs
\ni {\bf 4.5. Empirical constraints on the equation of state of
neutron star matter}

Current theoretical models do not predict the internal structure
of neutron stars and their equation of state in a unique way.
With increasing density a wide range of possibilities occurs as
far as relevant degrees of freedom are concerned. Uncertainties
regarding the equation of state and the internal structure are
discussed in Ref.[41]. 

The EOS of neutron star matter at high densities is not at
present constrained sufficiently by theory to allow conclusive
statements as to the internal  structure of neutron stars. In
view of weakness of theoretical constraints it is urgent to
empirically constrain the EOS.

\bs

\ni {\it 4.5.1. Laboratory experiments}

Scattering of heavy ions, which is the only laboratory way to
study properties of dense matter, does not probe directly the
EOS relevant to neutron stars. In nuclear collisions highly
excited hadronic matter is formed which decays quickly into
stable particles. One should perform extrapolations in order
to obtain the ground state properties of $\beta$-stable matter from
these data which would 
involve considerable uncertainty. However, many important
informations regarding the neutron star EOS can be inferred from
scattering data. In particular, interactions of hyperons formed
in nuclear collisions with nucleons in dense fireball can be
studied. Also, detection of quark-gluon plasma in heavy ion
collisions could give valuable informations about
energy density range in which one can expect deconfinement
transition in neutron star matter.

\bs

\ni {\it 4.5.2. Observations of neutron stars}

Discovery of a sub-millisecond pulsar would severely constrain
the EOS. Unfortunately, reported observations of 0.5 ms 
pulsar in the SN1987A remnant [42] turned out to be erroneous.
The fastest millisecond pulsar of period 1.56 ms does not
exclude any realistic EOS.

\bs 

\ni {\it 4.5.3. X-rays from rotation-powered pulsars}

Observations of thermal flux of photons from a pulsar, whose age
can be estimated, can provide information how fast  the
neutron star cools. Probing the cooling curve of neutron
stars is considered to be the most profitable method to learn
about the internal composition of neutron stars.  

Recent observations of thermal X-ray flux from four pulsars give
promise that in near future the cooling curve will be
empirically constrained. The main objective of these
observations is to discriminate between fast and slow cooling
mechanisms. 

Slow cooling proceeds mainly through the modified
URCA process,

$$ n+n \to n+p+e^-+{\bar \nu_e},~~~~~~~ n+p+e^- \to
n+n+\nu_e. $$

\ni It is the dominating  mechanism of standard cooling
scenario for neutron stars
whose proton content is below the critical value, $x<x_{URCA}
\approx 0.11$.

If the proton content exceeds the critical value, $x>x_{URCA}$, or
there exists kaon (or pion) condensate in the neutron star matter,
cooling proceeds through direct URCA process,

$$ n \to p+e^-+{\bar \nu_e},~~~~~~~ p+e^- \to n+\nu_e. $$

\ni This cooling mechanism is much faster, and, correspondingly,
the temperature of the neutron star is lower than for modified
URCA. 

Recent comparison of the X-ray luminosities of four pulsars [43]
is not conclusive, but the observational data are somewhat
closer to the standard cooling curve.

\bs

\ni {\it 4.5.4. Remnant of SN1987A}

Detection of the neutrino flux associated with optical observations
of the supernova SN1987A was the best confirmation of the theory of
neutron star formation in supernova Type II explosions. Present
observations of the light curve of the remnant of SN1987A do not
confirm existence of the neutron star. Continuously decreasing
luminosity of the remnant of SN1987A suggests that Crab-like
pulsar does not exist in the remnant. Also, no hot X-ray source
is observed. This lack of signature of the neutron star has led some
authors [44] to speculate that the neutron star formed initially
in SN1987A was in a metastable state and subsequently collapsed
to a black hole. Presence of the black hole in SN1987A would strongly
constrain the neutron star EOS.
 
Scenario of black hole formation in SN1987A is as follows. The
progenitor star of SN1987A is known to have mass
$18M_{\odot}<M<20M_{\odot}$. Evolutionary calculations show 
that this star developed an iron core of mass $\sim 1.6 M_{\odot}$
which collapsed to form hot neutron star. This neutron star
existed at least for $\sim 10s$, a period when the neutrino
emission took place. After radiating away the trapped neutrinos
the neutron star has lost stability and collapsed to black hole.

This scenario requires that the EOS has some unique features.
The maximum mass corresponding to the hot neutron star matter,
with trapped neutrinos, $M_{max}^{hot}$, has to be higher than the
mass $M_{ns}\sim 1.6 M_{\odot}$ of the
neutron star formed in the collapse of SN1987A,
$M_{ns}<M_{max}^{hot}$.  After emitting neutrinos
the neutron star looses stability which requires that
its mass is higher than the maximum mass of a neutron star
corresponding to cold neutron star matter, $M_{max}^{cold}<M_{ns}$.
The EOS with kaon condensation can meet these constraints [45].
The maximum mass for cold EOS is
$M_{max}^{cold}=1.5 M_{\odot}$.

The maximum mass of neutron star of $1.5 M_{\odot}$ explains in
a natural way the cutoff observed in measured masses of neutron
stars. Existence of such a limit is very surprising in view of
the fact that considerable amount of material, at least a few
tenths of solar mass, is expected to fall back onto newly formed
neutron star after the explosion. One would expect many heavier
neutron stars to be formed.

If the maximum mass of neutron star is $M_{max}=1.5 M_{\odot}$
then one can determine maximum mass of the progenitor star,
whose collapse can leave the neutron star remnant. For an
isolated star this mass is about $20 M_{\odot}$. Heavier stars
are expected to leave black hole remnants.

\bs
This work is partially supported by the Polish State Committee
for Scientific Research (KBN), grants 2 P03D 001 09 and 2 P03B
083 08. The author is grateful for inviting him to give these
lectures at the XXV Mazurian Lakes School of Physics.

\bs\bs

\ni {\bf References}

\ni~1.~~~ S. L. Shapiro and S. A. Teukolsky, {\it Black Holes, White
Dwarfs and Neutron Stars}, 

\ni~~~~~~Wiley, 1983.

\ni~2.~~~F. M. Walter and L. D. Metthews, Nature {\bf 389}, 358 (1997).

\ni~3.~~~F. Camilo, S. E. Thorsett, and S. R. Kulkarni, Astrophys. J. 

\ni~~~~~~{\bf 421}, L15 (1994).

\ni~4.~~~F. Haberl, L. Angelini, C. Motch, and N. E. White, astro-ph/9710138.

\ni~5.~~~L. Bildsten et al., astro-ph/9707125.

\ni~6.~~~R. A. Hulse and J. H. Taylor, Astrophys. J. {\bf 195},
L51 (1975).

\ni~7.~~~F. M. Walter, S. J. Wolk, and R. Neuh\"auser, Nature
{\bf 379}, 233 (1996).

\ni~8.~~~C. Alcock et al., Astrophys. J. {\bf 445}, 133 (1995).

\ni~9.~~~J. H. Taylor and J. M. Weisberg, Astrophys. J. {\bf
345}, 434 (1989).

\ni~10.~~A. Wolszczan, Nature {\bf 350}, 688 (1991).

\ni~11.~~S. E. Thorsett, Z. Arzoumanian, M. M. McKinnon and J. H.
Taylor, Astrophys. J. 

\ni~~~~~~{\bf 405}, L29 (1993).

\ni~12.~~W. T. S. Deich and S. R. Kulkarni, in {\it Compact Stars
in Binaries}, eds J. van Paradijs, 

\ni~~~~~~E. P. J. van den Heuvel and E.
Kuulkers eds., Dordrecht, Kluwer 1996.

\ni~13.~~Z. Arzoumanian, Ph. D. thesis, Princeton Univ. (1995).

\ni~14.~~D. J. Nice, R. W. Sayer and J. H. Taylor, Astrophys. J.
{\bf 466}, L87 (1996).

\ni~15.~~H. \"Ogelman, in {\it The Lives of Neutron Stars}, eds M. A.
Alpar, U. Kiziloglu and J. van 

\ni~~~~~~Paradijs eds., Dordrecht, Kluwer
1995. 

\ni~16.~~A. G. Lyne, R. S. Pritchard, F. Graham-Smith and F.
Camilo, Nature {\bf 381}, 

\ni~~~~~~497 (1996).

\ni~17.~~K. Makishima, in {\it The Structure and Evolution of
Neutron Stars}, eds D. Pines, 

\ni~~~~~~R. Tamagaki and S. Tsuruta, Addison-Wesley 1992.

\ni~18.~~A. Blaauw, Bull. Astron. Inst. Neth. {\bf 505}, 265 (1961). 

\ni~19.~~C. Fryer and V. Kalogera, astro-ph/9706031.

\ni~20.~~V. Kalogera and R. F. Webbink, astro-ph/9708223.

\ni~21.~~E.P.J. van den Heuvel, in {\it Neutron Stars: Theory
and Observation}, 

\ni~~~~~~eds J. Ventura and D. Pines, Kluwer 1991.

\ni~22.~~M. Tavani, in {\it Neutron Stars: Theory
and Observation}, eds J. Ventura and D. Pines, 

\ni~~~~~~Kluwer 1991.

\ni~23.~~G. Baym, in {\it Neutron Stars: Theory
and Observation}, eds J. Ventura and D. Pines, 

\ni~~~~~~Kluwer 1991.

\ni~24.~~J. M. Lattimer, in {\it The Structure and Evolution of
Neutron Stars}, eds D. Pines, 

\ni~~~~~~R. Tamagaki and S. Tsuruta, Addison-Wesley 1992.

\ni~25.~~D. Vautherin and D. M. Brink, Phys. Lett. {\bf 32B},
149 (1970).

\ni~26.~~W. D. Myers and W. J. Swiatecki, Nucl. Phys. {\bf A
601}, 141 (1996).

\ni~27.~~R. V. Reid, Ann. Phys. {\bf 50}, 411 (1968).

\ni~28.~~I. E. Lagaris and V. R. Pandharipande, Nucl. Phys. {\bf
A359}, 331, 349 (1981).

\ni~29.~~R. W. Wiringa, R. A. Smith and T. L. Ainsworth, Phys. Rev. C
{\bf 29}, 1207 (1984).

\ni~30.~~R. W. Wiringa, V. Fiks and A. Fabrocini, Phys. Rev. C
{\bf 38}, 1010 (1988).

\ni~31.~~M. Kutschera, Z. Phys. A {\bf 348}, 263 (1994).
 
\ni~32.~~M. Kutschera, Phys. Lett. {\bf B340}, 1 (1994).

\ni~33.~~M. Kutschera and W. W\'ojcik, Phys. Rev. C {\bf 47},
1077 (1993).

\ni~34.~~M. Kutschera and W. W\'ojcik, Acta Phys. Pol. {\bf
B21}, 823 (1990).

\ni~35.~~J. Bardeen and W. Shockley, Phys. Rev. {\bf 80}, 72 (1950).

\ni~36.~~M. Kutschera and W. W\'ojcik, Phys. Lett. {\bf B223}, 1
(1989).

\ni~37.~~M. Kutschera and W. W\'ojcik, Nucl. Phys. {\bf A581},
706 (1994).

\ni~38.~~M. Kutschera and W. W\'ojcik, Acta Phys. Pol. {\bf
A92}, 375 (1997).

\ni~39.~~M. Kutschera and W. W\'ojcik, Acta Phys. Pol. {\bf
B27}, 2227 (1996).

\ni~40.~~M. Kutschera and W. W\'ojcik, Acta Phys. Pol. {\bf
B23}, 947 (1992).

\ni~41.~~M. Kutschera, in {\it Solar Astrophysics, Structure of
Neutron Stars, Gamma Flares}, 

\ni~~~~~~ed. K. Grotowski, Polish Academy
of Arts and Sciences, Cracow, 1997.

\ni~42.~~C. Kristian, et al., Nature {\bf 338}, 234 (1989).

\ni~43.~~Ch. Schaab, F. Weber, M. K. Weigel and N. K.
Glendenning, astro-ph/9603142.

\ni~44.~~G. E. Brown, S. W. Bruenn and J. C. Wheeler, Comments
Astrophys. {\bf 16}, 153 (1992).

\ni~45.~~V. Thorsson, M. Prakash and J. M. Lattimer, Nucl. Phys.
{\bf A572}, 693 (1994).

\vfill
\break

\bigskip
\bigskip
\epsffile {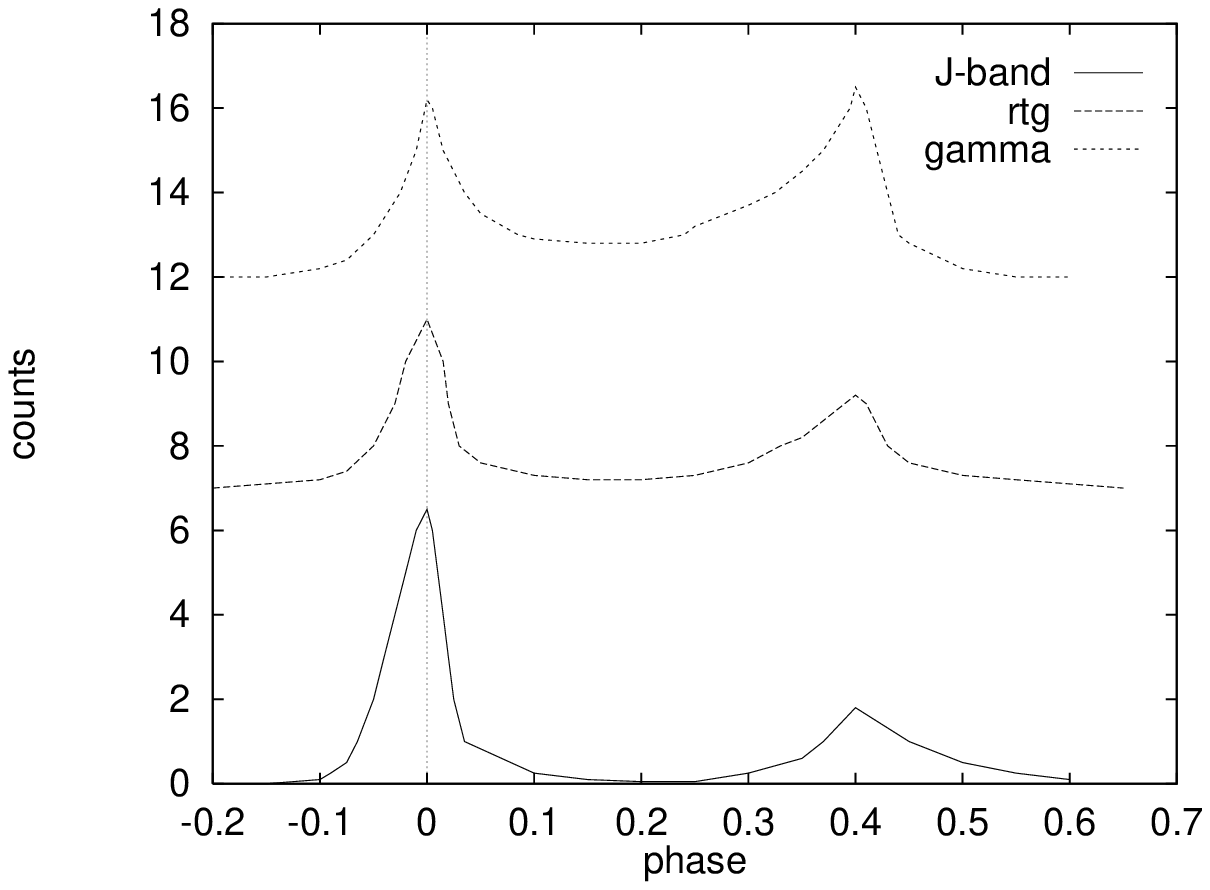}

\noindent Fig.1

\ni The pulse profile of the Crab pulsar in infrared J-band, X-rays and
$\gamma$-rays. Curves for different bands are shifted
vertically. Units on vertical axis are arbitrary.

\vfill
\break

\bigskip
\epsffile {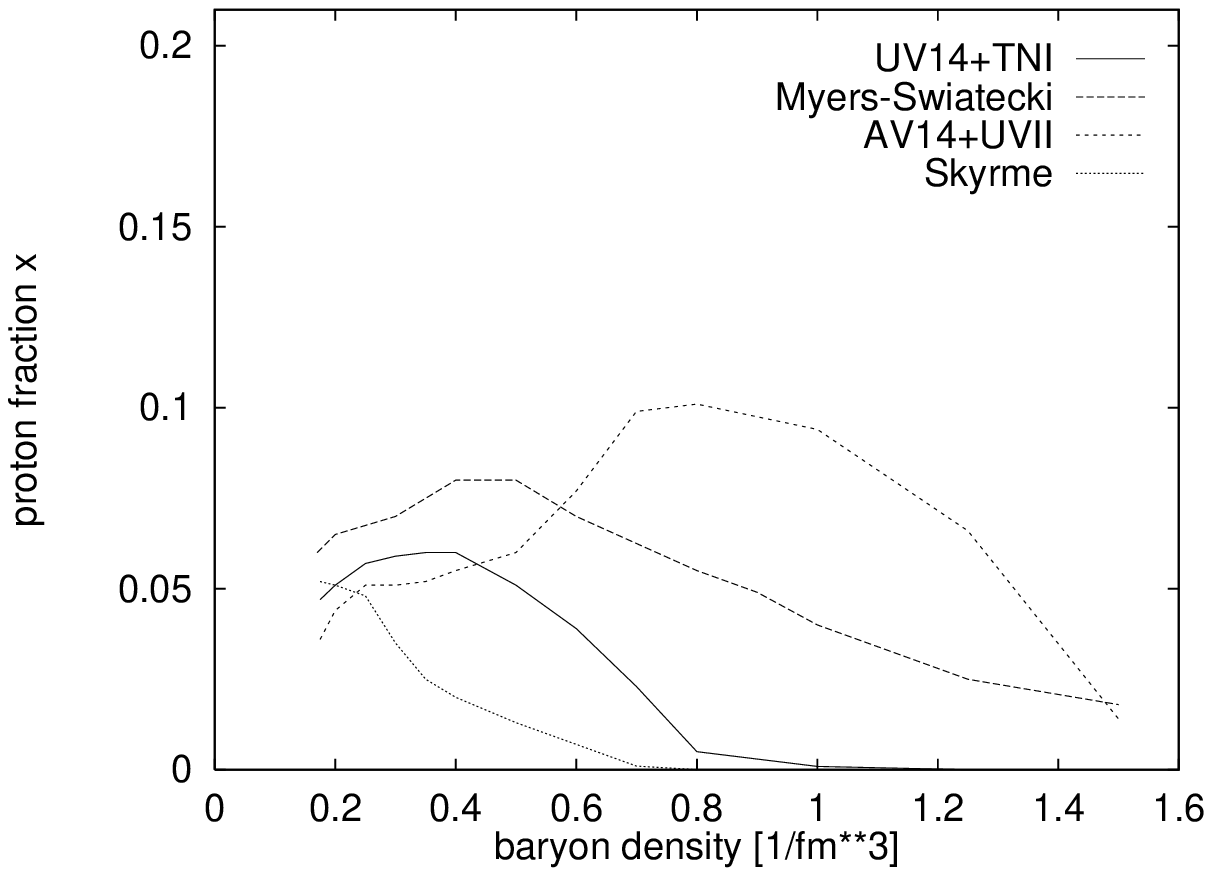}
\noindent Fig.2

\noindent The proton fraction $x$ as a function of density for realistic
models of neutron star matter based on Skyrme forces, Myers and
Swiatecki interactions, and on Urbana $v_{14}$ and Argonne
$v_{14}$ potentials.

\vfill
\break

\bigskip

\epsffile{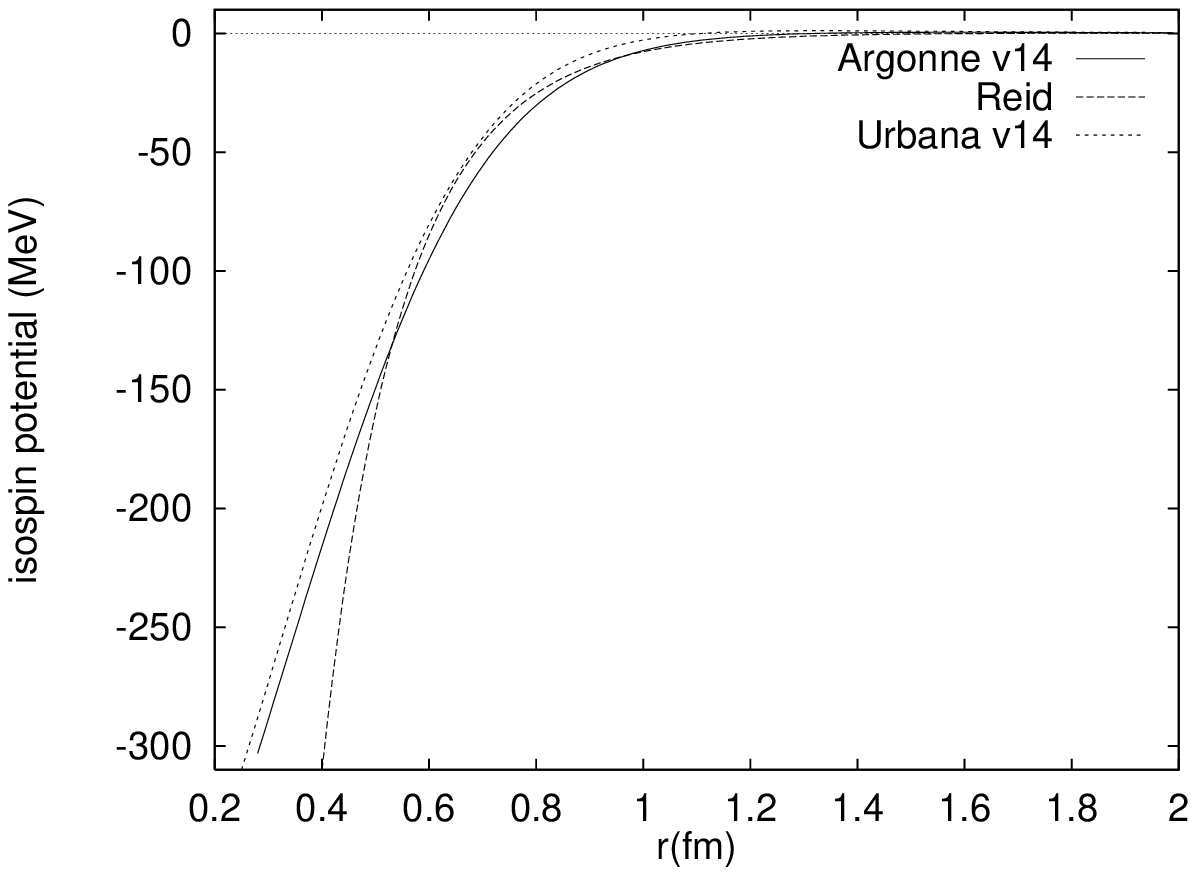}
\noindent Fig.3

\noindent The isospin potential $v_{\tau}$ of  Reid, Argonne $v_{14}$, and Urbana
$v_{14}$ nuclear potentials.

\vfill
\end